\documentclass[paper]{JHEP3}

\usepackage{epsfig,multicol,multirow}
\usepackage{amsmath,subfigure,latexsym,amssymb}

\def\lsi{\raise0.3ex\hbox{$<$\kern-0.75em\raise-1.1ex\hbox{$\sim$}}}
\def\gsi{\raise0.3ex\hbox{$>$\kern-0.75em\raise-1.1ex\hbox{$\sim$}}}

\newcommand\fverb{\setbox\pippobox=\hbox\bgroup\verb}
\newcommand\fverbdo{\egroup\medskip\noindent%
                        \fbox{\unhbox\pippobox}\ }
\newcommand\fverbit{\egroup\item[\fbox{\unhbox\pippobox}]}
\newbox\pippobox
\newcommand{\beq}{\begin{equation}}
\newcommand{\eeq}{\end{equation}}
\newcommand{\beqa}{\begin{eqnarray}}
\newcommand{\eeqa}{\end{eqnarray}}
\newcommand{\Z}{\mathbb{Z}}
\newcommand{\R}{\mathbb{R}}
\newcommand{\dd}{\mbox{d}}
\newcommand{\del}{\partial}
\newcommand {\tr}{{\rm tr\,}}

\newcommand {\defeq}{\stackrel{\rm def}{=}}

\preprint{
SAGA-HE-224 \\ 
KEK-TH-1059 \\
UTHEP-515}
\title{
The index of the overlap Dirac operator
on a discretized 2d non-commutative torus
}

\author{
Hajime Aoki${}^{a}$,
Jun Nishimura${}^{bc}$ and
Yoshiaki Susaki${}^{bd}$\\ 
\llap{$^a$}Department of Physics, Saga University, 
Saga 840-8502, Japan \\
\llap{$^b$}High Energy Accelerator Research Organization (KEK), \\
Tsukuba, Ibaraki 305-0801, Japan \\
\llap{$^c$}Department of Particle and Nuclear Physics,\\
Graduate University for Advanced Studies (SOKENDAI),\\
Tsukuba, Ibaraki 305-0801, Japan \\
\llap{$^d$}Graduate School of Pure and Applied Science,
University of Tsukuba,\\
Tsukuba, Ibaraki 305-8571, Japan\\
\email{haoki@cc.saga-u.ac.jp,
jnishi@post.kek.jp,
susaki@post.kek.jp}} 

\abstract{
The index,
which is given in terms of 
the number of zero modes of the Dirac operator
with definite chirality, 
plays a central role in various topological aspects of gauge theories.
We investigate its properties
in non-commutative geometry.
As a simple example,
we consider the U(1) gauge theory on a 
discretized 2d non-commutative 
torus, in which general classical solutions are known. 
For such backgrounds we calculate
the index of the overlap Dirac operator satisfying 
the Ginsparg-Wilson relation.
When the action is small,
the topological charge defined by a naive discretization
takes approximately integer values, and it agrees with the index
as suggested by the index theorem.
Under the same condition, the value of the 
index turns out to be a multiple of $N$, the size of the 2d lattice.
By interpolating the classical solutions, we construct explicit 
configurations, for which the index is of order 1, but
the action becomes of order $N$.
Our results suggest that
the probability of obtaining 
a non-zero index vanishes in the continuum limit,
unlike the corresponding results in the commutative space.
}


\begin{document}

\section{Introduction}

Non-commutative (NC) geometry \cite{Sny,Connes}
has been studied for quite a long time
as a simple modification of our notion of space-time
at small distances possibly due to effects of 
quantum gravity \cite{gravity}.
It has attracted much attention since it was shown to 
appear naturally from matrix models \cite{CDS,AIIKKT}
and string theories \cite{String}.
In particular, field theory on NC geometry has
a peculiar property known as the UV/IR mixing \cite{rf:MRS},
which may cause a drastic change of the long-distance physics
through quantum effects.
This phenomenon has been first discovered in perturbation theory, 
but it was shown to appear also in a fully nonperturbative
setup \cite{AMNS}.
A typical example is the spontaneous breaking of the
translational symmetry in NC scalar field theory, 
which was first conjectured 
from a self-consistent one-loop analysis \cite{GuSo}
and confirmed later on by Monte Carlo simulation 
\cite{Procs,AC,Bietenholz:2004xs}. (See also \cite{ChenWu,CZ}.)

The appearance of a new type of IR divergence due to the UV/IR mixing
spoils the perturbative renormalizability in general
\cite{Chepelev:1999tt},
and therefore, even the existence of a sensible field theory on a NC
geometry is {\em a priori} debatable.
In order to study such a nonperturbative issue,
one has to define a regularized field theory on NC geometry,
which is possible by using matrix models.
In the case of NC torus, for instance, 
the so-called twisted reduced model \cite{EK,GAO}
is interpreted
as a lattice formulation of NC field theories \cite{AMNS},
in which finite $N$ matrices are mapped
one-to-one onto fields on a periodic lattice.
The existence of a sensible continuum limit
and hence the nonperturbative renormalizability
have been shown by Monte Carlo simulations
in NC U(1) gauge theory in 2d \cite{2dU1} and 4d
\cite{4dU1}
as well as in NC scalar field theory 
in 3d \cite{Bietenholz:2004xs,Bietenholz:2004as}.

In the case of fuzzy sphere \cite{Madore}, finite $N$ matrices
are mapped one-to-one onto functions on the sphere
with a specific cutoff on the angular momentum.
The fuzzy sphere (or fuzzy manifolds 
\cite{Dolan:2003kq,O'Connor:2003aj} in general)
preserves the continuous
symmetry of the base manifold, which makes it an interesting 
candidate for a novel regularization of {\em commutative} field theories
alternative to the lattice \cite{Grosse:1995ar}.
It is also interesting to use fuzzy spheres in
the coset space dimensional reduction \cite{Aschieri:2003vy}.
Stability of fuzzy manifolds in
matrix models with the Chern-Simons term \cite{Myers:1999ps,0101102}
has been studied by Monte Carlo simulations
\cite{Azuma:2004zq,fuzzy-MC}.

One of the interesting features of NC field theories
is the appearance of a new type of topological objects,
which are referred to as NC solitons \cite{GMS},
NC monopoles, NC instantons, and fluxons \cite{fluxon}
in the literature.
They are constructed by using a projection operator,
and the matrices describing such configurations
are assumed to be infinite dimensional.
In finite NC geometries,
namely in the case where
field configurations are described by finite-dimensional matrices
and therefore regularized,
topological objects have been constructed 
by using the algebraic K-theory and projective modules 
\cite{non-trivial_config,
Balachandran:2003ay,AIN3}.

Dynamical aspects of these topological objects are
important in particular
in the realization of a 4d chiral gauge theory
in the context of string theory compactification,
which requires a nontrivial index in the compactified 
dimensions (See, for instance, Chapter 14 of ref.\ \cite{GSW}.)
Ultimately we hope to realize such a scenario {\em dynamically},
for instance, in the IIB matrix model \cite{9612115},
in which the dynamical generation of {\em four-dimensional} space-time
\cite{Aoki:1998vn,spaceiib,%
gaussian}
as well as the gauge group \cite{Iso:1999xs,0504217}
has been studied intensively.

Extending the notion of the index to {\em finite} NC geometry
is a nontrivial
issue due to the doubling problem of the naive Dirac action.
The same problem occurs also in the ordinary lattice gauge theory.
%
There one can add the Wilson term to the Dirac action 
to remove the species doublers
in the continuum limit at the expense of the explicit
breaking of chiral symmetry.
This had been a notorious problem in lattice gauge theory
as manifested by the no-go theorem \cite{Nielsen:1980rz}.
It was found, however, that 
by adopting a Dirac operator which satisfies 
the Ginsparg-Wilson relation \cite{GinspargWilson},
a modified chiral symmetry,
which becomes the usual one in the continuum limit,
can be exactly preserved on the lattice \cite{Luscher,Nieder}.
The Dirac operator has exact zero modes with definite chirality
for topologically nontrivial gauge configurations, and therefore
one can define the index unambiguously 
\cite{Narayanan:1994gw,Hasenfratzindex,Luscher}.
A concrete example of such an operator with desirable
properties in the continuum limit is given by the so-called
overlap\footnote{Historically, 
the overlap formalism \cite{Narayanan:1994gw},
from which one can actually derive the overlap Dirac operator 
\cite{Neuberger}, has been established before the rediscovery
of the Ginsparg-Wilson relation.}
Dirac operator \cite{Neuberger}. 
The index theorem for the overlap Dirac operator 
is studied numerically in refs.\
\cite{Narayanan:1994gw,index_commutative}.
The successful results obtained in these tests 
may be understood from the analytical 
work \cite{Adams} (See also ref.\ \cite{Adams:2003hy} for an extension.),
in which the usual expression for the topological charge 
in the continuum has been derived
from the index of the overlap Dirac operator {\em nonperturbatively}.
As emphasized in refs.\ \cite{Adams},
the derivation of the correct axial anomaly 
\cite{axial-anomaly-comm},
which uses the perturbative expansion with respect to the gauge field,
is not sufficient for demonstrating
the index theorem for topologically nontrivial gauge configurations.

In the past several years, the ideas developed in lattice gauge theory
have been successfully extended to NC geometry.
In the case of NC torus, the overlap Dirac operator has been
introduced in ref.\ \cite{Nishimura:2001dq}, and 
it was used to define a NC chiral gauge theory
with manifest star-gauge invariance.
For general NC manifolds,
a prescription to define 
the Ginsparg-Wilson Dirac operator and its index
has been provided in ref.\ \cite{AIN2},
and the fuzzy sphere was considered as a concrete 
example\footnote{The Ginsparg-Wilson Dirac operator 
for vanishing gauge field 
was constructed earlier in refs.\ \cite{balagovi}.}.
The Ginsparg-Wilson algebra
for the fuzzy sphere 
has been studied in detail in each topological 
sector \cite{Balachandran:2003ay}.
%
In ref.\ \cite{isonagao}
the overlap Dirac operator on the NC torus 
\cite{Nishimura:2001dq}
was derived also from this general prescription \cite{AIN2},
and the axial anomaly has been calculated in the continuum limit.

In an attempt to construct a topologically nontrivial configuration
on the fuzzy sphere, an analogue of the 't Hooft-Polyakov monopole
was obtained \cite{Balachandran:2003ay,AIN3}.
Although the index defined through the Ginsparg-Wilson Dirac operator
vanishes for these configurations, one can 
make it non-zero
by inserting a projection operator, which
picks up the unbroken U(1) component of the SU(2) gauge group.
In fact the 't Hooft-Polyakov monopole configurations are precisely
the meta-stable states observed in Monte Carlo simulations \cite{Azuma:2004zq}
taking the two coincident fuzzy spheres as the initial configuration,
which eventually decays into a single fuzzy sphere.
In ref.\ \cite{AIMN}
this instability was studied analytically by
the one-loop calculation of free energy
around the 't Hooft-Polyakov monopole 
configurations,
and it was interpreted as
the dynamical generation of a nontrivial index,
which may be used for the realization of a chiral fermion in our space-time.
%

The primary aim of the present work is to investigate the
properties of the index in finite NC geometry,
taking the 2d U(1) gauge theory on a discretized NC torus as a 
simple example,
which is studied extensively in the literature
both numerically \cite{2dU1}
and analytically \cite{Mafia,Paniak:2002fi,
Griguolo:2003kq}.
In particular, ref.\ \cite{Griguolo:2003kq} presents
general classical solutions carrying the topological 
charge.
We compute the index defined through the overlap 
Dirac operator
for these classical solutions.
%
%
The topological charge defined naively on the
discretized NC torus is not an integer in general,
although the index is.
We observe, however, that when the action is small,
the topological charge is close to an integer,
and it agrees with the index as suggested by the index theorem.
In fact, under the same condition,
the index turns out to be
a multiple of $N$, the linear size of the 2d lattice.
%
By interpolating the classical solutions, we construct explicit 
configurations\footnote{While we were preparing this article, 
we received a preprint
\cite{Nagao:2005st}, in which a gauge configuration
with non-zero index
was
found numerically in the same model at small $N$.
}
for which the index is of order 1, but
the action becomes of order $N$.
Our results suggest that the probability of obtaining 
a non-zero index vanishes in the continuum limit, 
which is
consistent with the instanton calculus in the continuum theory
\cite{Paniak:2002fi}.
%


The rest of this paper is organized as follows.
In section \ref{model} 
we provide some generalities concerning
a matrix model formulation of 
gauge theories on a discretized NC torus and 
define the index of the overlap 
Dirac operator.
In section \ref{section:two-dim-case}
we focus on the two-dimensional case,
and discuss the classical solutions and
the topological charge.
In section \ref{section:results} we examine whether
the index theorem holds for the classical solutions.
In section \ref{section:interpolating} we construct explicit 
configurations with the index of order 1 by interpolating
the classical solutions, and study their properties.
In section \ref{section:commutative} we review some known results
in the commutative case, and discuss their relationship to 
our results.
Section \ref{summary} is devoted to a summary and discussions.


\section{Generalities}
\label{model}

\subsection{gauge theory on a discretized NC torus}
\label{2dNCgauge}

The U(1) gauge theory on a NC space is given by the action
\beqa
\label{cont-action}
S_{\rm cont}&=& \frac{1}{g^2}\int
\dd ^ d x ~ \frac{1}{4}\Bigl(F_{\mu\nu}(x) \star F_{\mu\nu}(x)\Bigr) \ , \\
F_{\mu\nu}(x)&=& \del _\mu A_\nu(x) - \del _\nu A_\mu(x)
+ i \, \left\{ A_\mu(x) \star A_\nu(x) - A_\nu(x) \star A_\mu(x)\right\} \ ,
\label{def-F}
\eeqa
where the star-product is defined by
\beq
\varphi_1(x)\star \varphi_2(x)=
\left.\exp\left(\frac{i}{2}\Theta_{\mu\nu}
\frac{\del}{\del x_\mu}\frac{\del}{\del y_\mu}\right)
\varphi_1(x)\varphi_2(y)\right|_{x=y} \ .
\label{def-starprod}
\eeq
Note that the star-product is associative but non-commutative.
This non-commutativity may be attributed to that of space-time
since
\beq
x_\mu \star x_\nu - x_\nu \star x_\mu = i \Theta_{\mu\nu} \ .
\eeq
The action (\ref{cont-action})
is invariant under a star-gauge transformation
\beq
A_\mu(x) \mapsto 
g(x)\star A_\mu(x) \star g^\ast (x) 
- i g(x)\star \del_\mu g^\ast (x) \ ,
\eeq 
where $g(x)$ obeys
the star-unitarity condition
\beq
g(x) \star g(x)^{*} = 
 g(x)^{*} \star g(x) = 1
\label{star-unitary}
\eeq
instead of $|g(x)| = 1$.

When we discretize the space, the consistency with the NC algebra
inevitably requires the space to be compactified 
in a specific way \cite{AMNS}.
Thus we obtain a theory on a periodic $L^d$ lattice
with the action
\beq
S_{\rm lat}= - \beta \sum_{x} \sum_{\mu \neq \nu}
U_\mu (x) \star U_\nu (x + a \hat{\mu}) \star
U_\mu (x + a \hat{\nu})^{*} \star U_\nu (x)^{*}  \ ,
\label{lat-action}
\eeq
where the link variables $U_\mu(x)$ are star-unitary; i.e.,
\beq
U_\mu(x) \star U_\mu(x)^{*} = 
U_\mu(x)^{*} \star U_\mu(x) = 1  \ .
\label{star-unitary2}
\eeq
We use the standard notation in lattice gauge theory,
where $\hat{\mu}$ represents a unit vector in the $\mu$ direction,
and $a$ represents the lattice spacing.
The star-product on the lattice can be obtained by
rewriting (\ref{def-starprod}) in terms of Fourier modes
and restricting the momentum to be the one allowed on the lattice.
As in the commutative space,
one obtains the continuum action (\ref{cont-action})
from (\ref{lat-action}) in the $a \rightarrow 0$ limit
with the identification $\beta = \frac{1}{2 a^2 g^2}$ and
\beq
U_\mu(x) = {\cal P} \exp_{\star} 
\left( i \int _x ^{x+a \hat{\mu}} dz A_\mu(z) \right) \ .
\eeq

\subsection{the 
overlap
Dirac operator and its index}

In this section we define 
the overlap 
Dirac operator and its index
for a gauge configuration on the discretized NC torus
\cite{Nishimura:2001dq,AIN2,isonagao}.
All the formulae have the same form as in the usual lattice gauge theory
except for the use of the star product.

We consider a Dirac operator $D$ satisfying 
the Ginsparg-Wilson relation \cite{GinspargWilson}
\beq
\gamma_5 D + D\gamma_5 = a \, D\gamma_5 D \ .
\label{GWrel}
\eeq
Assuming the $\gamma_5$-hermiticity $D^\dagger =\gamma_5 D\gamma_5$,
we can define a hermitian operator $\hat\gamma_5$ by
\beq
{\hat \gamma_5}=\gamma_5\left(1- a D\right) \ ,
\eeq
which may be solved for $D$ as 
$D=\frac{1}{a} (1 - \gamma_5 \hat{\gamma}_5)$.
Then the Ginsparg-Wilson relation (\ref{GWrel}) is equivalent to
requiring $\hat\gamma_5$ to be unitary.
The overlap Dirac operator 
corresponds to taking $\hat\gamma_5$ to be \cite{Neuberger}
\beqa
\hat\gamma_5 &=& \frac{H}{\sqrt{H^2}} \ , \\
H &=& \gamma_5 \left(1-aD_{\rm W}\right) \ ,
\label{H-def}
\eeqa
where $D_{\rm W}$ is the Wilson-Dirac operator
\beq
D_{\rm W}=\frac{1}{2}\sum_{\mu=1}^d
\left\{\gamma_\mu\left(\nabla_\mu^* 
+\nabla_\mu \right) - a \nabla_\mu^* \nabla_\mu \right\}
\label{def-Wilson-Dirac}
\eeq
with $\nabla_\mu$ ($\nabla_\mu^*$)
being the covariant forward (backward) difference operator defined by
\beqa
\nabla_\mu \Psi(x)&=&
\frac{1}{a}\left[U_\mu(x)\star \Psi(x+a{\hat \mu})-\Psi(x)  \right] \ , \\
\nabla_\mu^* \Psi(x)&=&
\frac{1}{a}\left[\Psi(x) - U_\mu(x-a{\hat \mu})^\dagger 
\star \Psi(x-a{\hat \mu}) \right] \ .
\eeqa

Since the Ginsparg-Wilson relation (\ref{GWrel}) 
can be rewritten as
\beq
\gamma_5 D + D {\hat \gamma_5}=0 \ ,
\label{GWrelation}
\eeq
the lattice action
\beq
S=a^d \sum_x {\bar \Psi}(x) \star D\Psi(x)\\
\eeq
has the exact lattice chiral symmetry \cite{Luscher,Nieder}
\beq
\Psi(x) \mapsto {\rm e}^{i\alpha {\hat \gamma_5}} \, \Psi(x) \ ,
~~~~~ 
{\bar \Psi}(x)  \mapsto{\bar \Psi}(x)
\, {\rm e}^{i\alpha \gamma_5} \ .
\eeq
Note also that the space of zero modes of the Dirac operator $D$
is invariant under $\gamma_5$, which means that
one can define the index of $D$ unambiguously by
$\nu \equiv n_+ - n_-$,
where $n_{\pm}$ is the number of zero modes
with the chirality $\pm 1$.
It turns out that \cite{Narayanan:1994gw,Hasenfratzindex,Luscher}
\beq
\nu =\frac{1}{2} {\cal T}r ( \gamma_5 + \hat\gamma_5) 
= \frac{1}{2} {\cal T}r \frac{H}{\sqrt{H^2}} \ , 
\label{indexD-H}
\eeq
where ${\cal T}r$ represents a trace over the space
of the Dirac spinor field on the lattice.


\subsection{matrix formulation}

So far we have been using a formulation 
of NC geometry, in which the non-commutativity of the space-time 
is encoded in the star-product.
In fact it is much more convenient for our purpose to use
an equivalent formulation \cite{Gonzalez-Arroyo:1983ac}, 
in which one maps functions on a NC space to operators
so that the star-product becomes nothing but 
the usual operator product, which is non-commutative.
In particular, the coordinate operators $\hat{x}_\mu$
satisfy the commutation relation
$[\hat{x}_\mu , \hat{x}_\nu] = i \Theta_{\mu\nu}$.
In the discrete version,
one maps a field $\varphi (x)$ on the $L^d$ lattice 
onto a $N \times N$ matrix $\Phi$, where $N^2 = L^d$ in order
to match the degrees of freedom.
This map yields the following correspondence
\beqa
\varphi_1 (x) \star \varphi_2 (x) &\Leftrightarrow& 
\Phi_1 \Phi_2  \ , \\
\varphi (x+a\hat{\mu}) &\Leftrightarrow& 
\Gamma_\mu \Phi \Gamma_\mu^\dag \ , \\
\frac{1}{L^d} \sum_{x} \varphi (x) &\Leftrightarrow& 
\frac{1}{N}\tr \Phi \ .
\eeqa
The SU($N$) matrices $\Gamma_\mu$ ($\mu = 1, \cdots , d$)
represent a shift on the matrix side, and they
satisfy the 't Hooft-Weyl algebra
\beq
\label{tH-W-alg}
\Gamma_\mu \Gamma_\nu = {\cal  Z}_{\mu\nu}
\Gamma_\nu \Gamma_\mu   \ ,
\eeq
where ${\cal Z}_{\mu\nu}={\cal  Z}_{\nu\mu}^*$ is a
phase factor.
An explicit representation of $\Gamma_\mu$
in the $d=2$ case shall be given
in section \ref{section:explicit2d}.

%

Using the map,
one can reformulate the lattice theory (\ref{lat-action})
in terms of matrices.
The star-unitarity condition (\ref{star-unitary2}) on
the link variables $U_\mu (x)$
simply implies that the corresponding matrix $\hat{U}_\mu$
should be unitary.
The action (\ref{lat-action}) can be written as 
\beqa
S &=& - N \beta \, \sum_{\mu \ne \nu} 
\tr ~\Bigl\{\hat{U}_\mu\, (\Gamma_\mu 
\hat{U}_\nu \Gamma_\mu^\dag)\,
(\Gamma_\nu \hat{U}_\mu^\dag \Gamma_\nu^\dag)
\, \hat{U}_\nu^\dag \Bigr\}  + 2 \beta N^2 \\
&=& - N \beta \, \sum_{\mu \ne \nu} 
{\cal  Z}_{\nu\mu}
\tr ~\Bigl(V_\mu\,V_\nu\,V_\mu^\dag\,V_\nu^\dag\Bigr) 
 + 2 \beta N^2  \ ,
\label{TEK-action}
\eeqa
where $V_\mu \equiv \hat{U}_\mu \Gamma_\mu$ is a U($N$) matrix.
This is nothing but the twisted Eguchi-Kawai (TEK) model \cite{GAO},
which appeared in history as a matrix model equivalent to the 
large $N$ gauge theory \cite{EK}.
In fact we have added 
the constant term $2\beta N^2$ to
what we would obtain from (\ref{lat-action})
in order to make the absolute minimum of the action zero.
We use this convention in the rest of this paper.

The index can be calculated using eq.\ (\ref{indexD-H}),
where $H$ is defined by eq.\ (\ref{H-def}).
The only thing to note in transcription into matrices
is that the covariant forward and backward difference operators 
$\nabla_\mu$ and $\nabla_\mu^*$, which appear in the definition 
of the Wilson-Dirac operator (\ref{def-Wilson-Dirac}), should now be
defined as
\beqa
\nabla_\mu \Psi&=&
\frac{1}{a}\left[\hat{U}_\mu (\Gamma_\mu \Psi \Gamma_\mu^\dag)
- \Psi  \right]
= \frac{1}{a}\left[V_\mu \Psi \Gamma_\mu ^\dag
- \Psi  \right] \ , \\
\nabla_\mu^* \Psi &=&
\frac{1}{a}\left[\Psi - (\Gamma_\mu^\dagger \hat{U}_\mu^\dagger 
\Gamma_\mu) (\Gamma_\mu^\dagger \Psi  \Gamma_\mu) \right] 
=\frac{1}{a}\left[\Psi - V_\mu ^\dagger 
\Psi  \Gamma_\mu \right]  \  .
\label{def-cov-shift}
\eeqa
The index is simply given by half the difference between the
numbers of positive and negative eigenvalues of 
the hermitian matrix $H$.
In this calculation we may simply set $a=1$,
since the lattice spacing $a$ appearing in the definition
of the index actually cancels out as it should.
The computational effort for calculating
the index is of order $N^6$,
since we have to diagonalize the $2N^2 \times 2N^2$
hermitian matrix $H$.

\section{Two-dimensional case}
\label{section:two-dim-case}

In this section we focus on the two-dimensional case,
and discuss the classical solutions and
the topological charge.

\subsection{explicit representation}
\label{section:explicit2d}

An explicit form of the map between fields and matrices
in the two-dimensional case
is given, for instance, in ref.\ \cite{Nishimura:2001dq},
where the twist in eq.\ (\ref{tH-W-alg}) is given by
\beq
{\cal  Z}_{12}=\exp\left(2 \pi i \frac{M}{N} \right)  \ ,
\quad \quad \quad  M = \frac{N+1}{2} 
\label{def-twist}
\eeq
with $N$ being an odd integer.
The algebra (\ref{tH-W-alg}) can be realized by
\beq
\Gamma_1 = {\cal P}_N \ , \quad \Gamma_2 = ({\cal Q}_N)^M \ ,
\eeq
where we have defined the SU$(n)$ matrices
\beq
{\cal P}_n=
\begin{pmatrix}
0&1& & &0\cr &0&1& & \cr& 
&\ddots&\ddots& \cr& & &\ddots&1\cr 1& &
& &0\cr
\end{pmatrix}  
~~~~~~,~~~~~~
{\cal Q}_n=
\begin{pmatrix}
1& & & & \cr & e^{2\pi i/n}& & & \cr&
& e^{4\pi i/n}& & \cr& & &\ddots& \cr & & & & e^{2\pi i(n-1)/n}\cr
\end{pmatrix}  
\label{clockshift}
\eeq
obeying ${\cal P}_n {\cal Q}_n
=e^{2\pi i/n}\, {\cal Q}_n {\cal P}_n$ for later convenience. 
For this particular construction, which we are going to use
throughout this paper, it turns out that
the NC tensor, which appears in the star-product, is given by
\beqa
\Theta_{\mu\nu} &=& \vartheta \, \epsilon_{\mu\nu}  \ , \\
\vartheta &=&\frac{1}{\pi} N a^2  \ .
\label{theta-def}
\eeqa
Note that the linear size of the torus $\ell = N a$ goes to $\infty$
in the continuum limit $a \rightarrow 0$ fixing $\vartheta$.
A finite torus can be obtained 
by other constructions given in the first paper of refs.\ \cite{AMNS} 
and ref.\ \cite{Griguolo:2003kq}, which are mutually equivalent.

\subsection{definition of the topological charge}

Let us define the topological charge for
a gauge configuration on the discretized 2d torus.
In the language of fields, we define the topological charge as
\beq
Q = \frac{1}{4 \pi i}
\sum_{x} \sum_{\mu \nu} \epsilon_{\mu\nu}
U_\mu (x) \star U_\nu (x + a \hat{\mu}) \star
U_\mu (x + a \hat{\nu})^{*} \star U_\nu (x)^{*}  \ ,
\label{def-q}
\eeq
which is obtained as a {\em naive} discretization 
of
the topological charge in 2d gauge theory defined in the continuum
as
\beq
Q = \frac{1}{4 \pi} \int d^2 x \, 
\epsilon_{\mu\nu} F_{\mu\nu} \ .
\eeq

By using the map between fields and matrices,
the topological charge (\ref{def-q}) can be represented
in terms of matrices as
\beqa
Q&=& \frac{1}{4 \pi i} N \, \sum_{\mu \nu} \epsilon_{\mu\nu}
\tr ~\Bigl\{ \hat{U}_\mu\, (\Gamma_\mu 
\hat{U}_\nu \Gamma_\mu^\dag ) \,
(\Gamma_\nu \hat{U}_\mu^\dag \Gamma_\nu^\dag)
\, \hat{U}_\nu^\dag \Bigr) \\
&=& \frac{1}{4 \pi i} N \, \sum_{\mu \nu} \epsilon_{\mu\nu}
{\cal  Z}_{\nu\mu}
\tr ~\Bigl(V_\mu\,V_\nu\,V_\mu^\dag\,V_\nu^\dag\Bigr) \ .
\label{def-q-mat}
\eeqa

\subsection{classical solutions}

The classical equation of motion can be obtained from 
the action (\ref{TEK-action}) as
\beq
V_\mu ^\dag (W - W^\dag ) V_\mu = W - W^\dag  \ ,
\eeq
where the unitary matrix $W$ is defined by
\beq
W = {\cal  Z}_{\nu\mu} V_\mu\,V_\nu\,V_\mu^\dag\,V_\nu^\dag \ .
\eeq
The general solutions\footnote{In fact there is 
another type of solutions,
which we do not consider in this paper
since they do not have finite action 
in the continuum limit \cite{Griguolo:2003kq}.
}
to this equation can be brought
into a block-diagonal form \cite{Griguolo:2003kq}
\begin{eqnarray}
V_\mu = 
\begin{pmatrix}
\Gamma_{\mu} ^{(1)} & & & \cr 
& \Gamma_{\mu} ^{(2)} & & \cr
& & \ddots & \cr 
& & & \Gamma_{\mu} ^{(k)} \cr  
\end{pmatrix}  
\label{general_sol}
\end{eqnarray}
by an appropriate SU($N$) transformation, where $\Gamma_{\mu} ^{(j)}$
are $n_j \times n_j$ unitary matrices satisfying the 't Hooft-Weyl algebra
\beqa
\Gamma_\mu ^{(j)} \Gamma_\nu ^{(j)} &=& Z_{\mu\nu}^{(j)}
\Gamma_\nu ^{(j)} \Gamma_\mu ^{(j)}  \  ,  \\
Z_{12}^{(j)} &=& Z_{21}^{(j)*} = \exp 
\left( 2\pi i \frac{ m_j }{n_j}  \right) \ .
\eeqa
An explicit representation \cite{vanBaal:1983eq}
is given, for instance, by 
\beq
\Gamma_1 ^{(j)}  = {\cal P}_{n_j} \ , 
\quad \Gamma_2  ^{(j)} = ({\cal Q}_{n_j})^{m_j} \ .
\label{explicit-Gamma}
\eeq
For each solution, the action and the topological charge
can be easily evaluated as
\beqa
\label{action-classical}
S &=& 4 N \beta
\sum_j n_j \sin ^2 \left\{ \pi \left( 
\frac{m_j}{n_j} - \frac{M}{N}
\right) \right\}   \  ,  \\
Q &=& \frac{N}{2\pi} \sum_j n_j \sin \left\{ 2 \pi  \left( 
\frac{m_j}{n_j} - \frac{M}{N} \right) \right\} \ .
\label{q-classical}
\eeqa
Note that the topological charge $Q$ is not an integer in general.
If we require the action to be less than of order $N$, however,
the argument of the sine has to vanish in the large $N$ limit
for all $j$ .
In that case the topological charge approaches an integer
\beq
Q \simeq N \left( \sum_j m_j  - M \right) \ ,
\label{q-integer}
\eeq
which is actually a multiple of $N$.

    \FIGURE{
    \epsfig{file=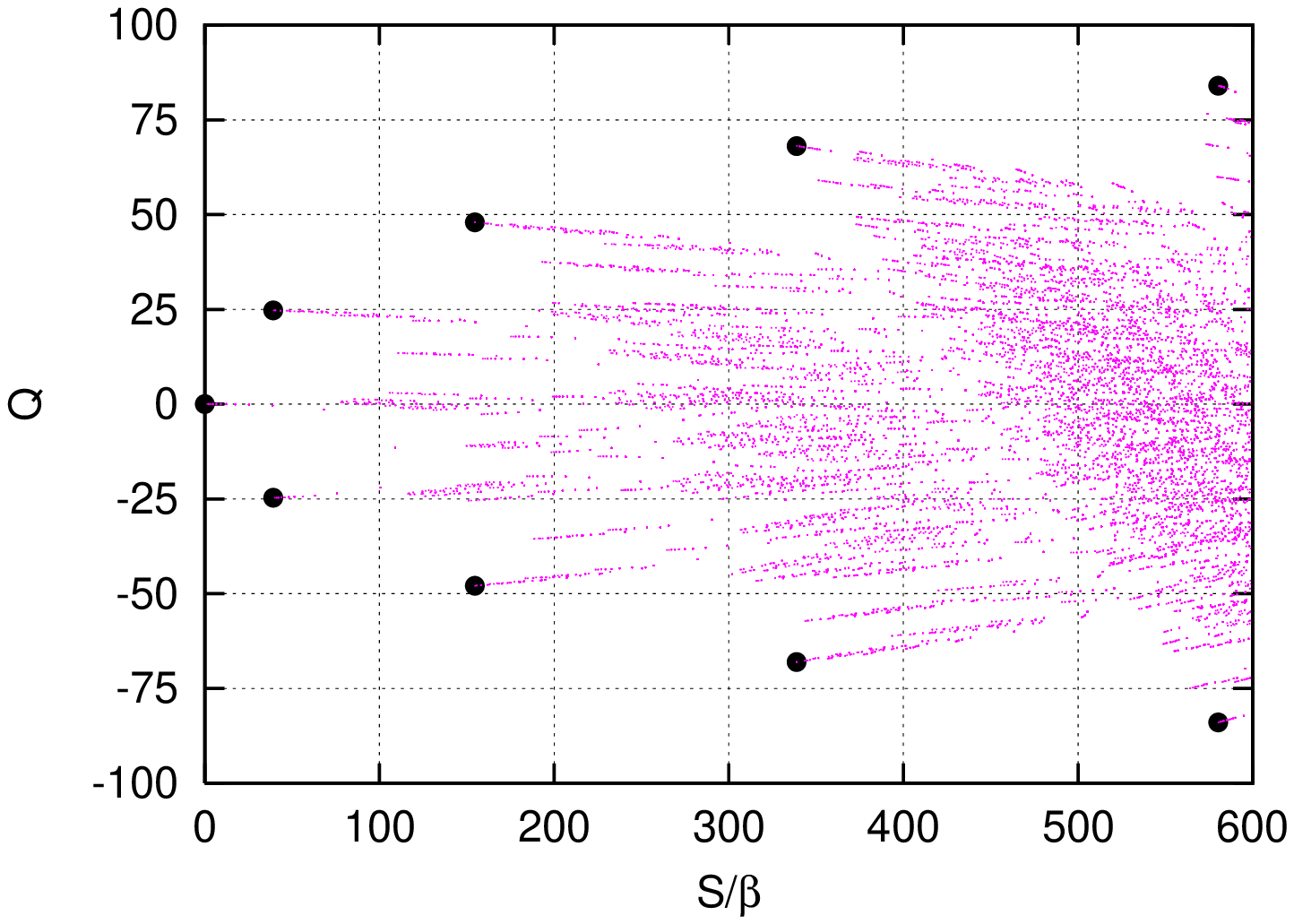,%
angle=270,width=7.4cm}
    \epsfig{file=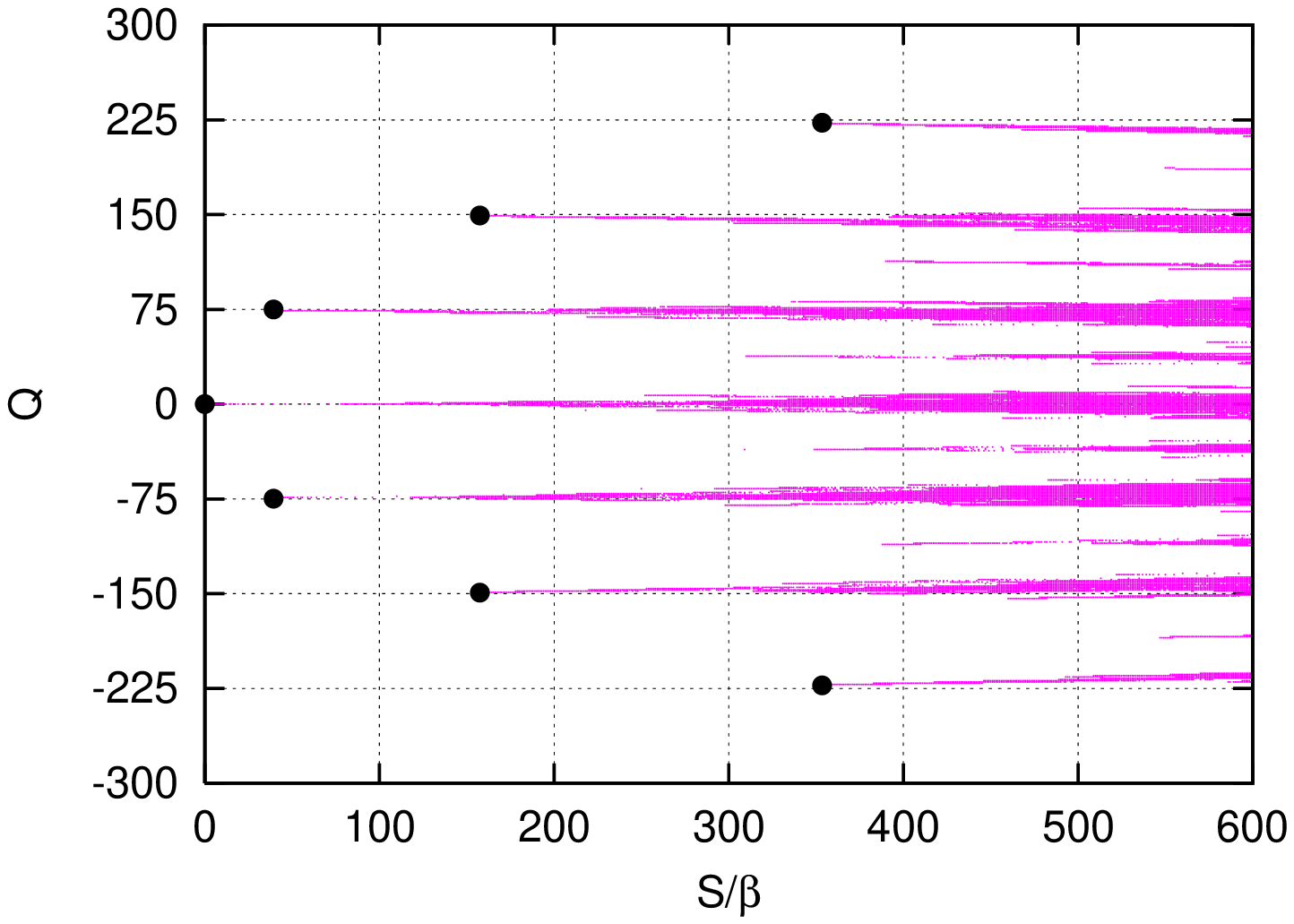,%
angle=270,width=7.4cm}
\caption{A scatter plot of the action $S/\beta$ (x-axis) and
the topological charge $Q$ (y-axis) for classical 
solutions at $N=25$ (left) and $N=75$ (right).
The closed circles represent the solutions for $k=1$,
which gives (\ref{action-singleblock}).
}
\label{scatter-q-S}
}

\section{The index theorem for the classical solutions}
\label{section:results}

The index theorem \cite{Atiyah:1971rm} relates 
the topological charge of an arbitrary gauge configuration
to the index of the Dirac operator on that background.
A proof of the index theorem
in noncommutative $\R ^ d$ is given 
in ref.\ \cite{Kim:2002qm}\footnote{In 
ref.\ \cite{Kim:2002qm}
an explicit ADHM construction of the fermionic zero modes 
in the multi-instanton backgrounds was also performed,
and the number of zero modes agreed with the index theorem.
See also ref.\ \cite{Sako:2002mq} for a related work.}.
However, as in the commutative case, the formulation of the index theorem
becomes nontrivial in the discretized setup.
Here we address this issue for the classical solutions
reviewed in the previous section.

Let us first look at the spectrum of the topological charge
for the classical solutions.
In figure \ref{scatter-q-S} we present 
a scatter plot of the action $S/\beta$ (x-axis) and
the topological charge $Q$ (y-axis) for the classical 
solutions at $N=25$ (left) and $N=75$ (right).
We plot all the solutions in the displayed range
without any restrictions\footnote{For $N=75$ this 
calculation was quite time-consuming 
because there are so many classical solutions. 
However, the figure looks almost the same
even if we restrict the number of blocks 
$k$ in eq.\ (\ref{general_sol}) to be e.g., $k\le 10$.}.
%
%
%
%
%
We observe the accumulation of 
solutions with the topological charge close to
a multiple integer of $N$.
The region of action, for which we obtain
only solutions with the 
topological charge close to a multiple integer of $N$,
extends with $N$.
This agrees with the argument that led to (\ref{q-integer})
in the previous section.

The minimum action in each topological sector
is achieved by the $k=1$ case,
for which eqs.\ (\ref{action-classical}) 
and (\ref{q-classical}) lead to
\beq
S \simeq 4 \pi^2 \beta  \left( \frac{Q}{N} \right)^2
\label{action-singleblock}
\eeq
at large $N$.
Note, however, that there are many solutions with $k>1$
which have an action very close to (\ref{action-singleblock}). 
For solutions with larger action, on the other hand,
the topological charge takes quite arbitrary values as expected.

    \FIGURE{
    \epsfig{file=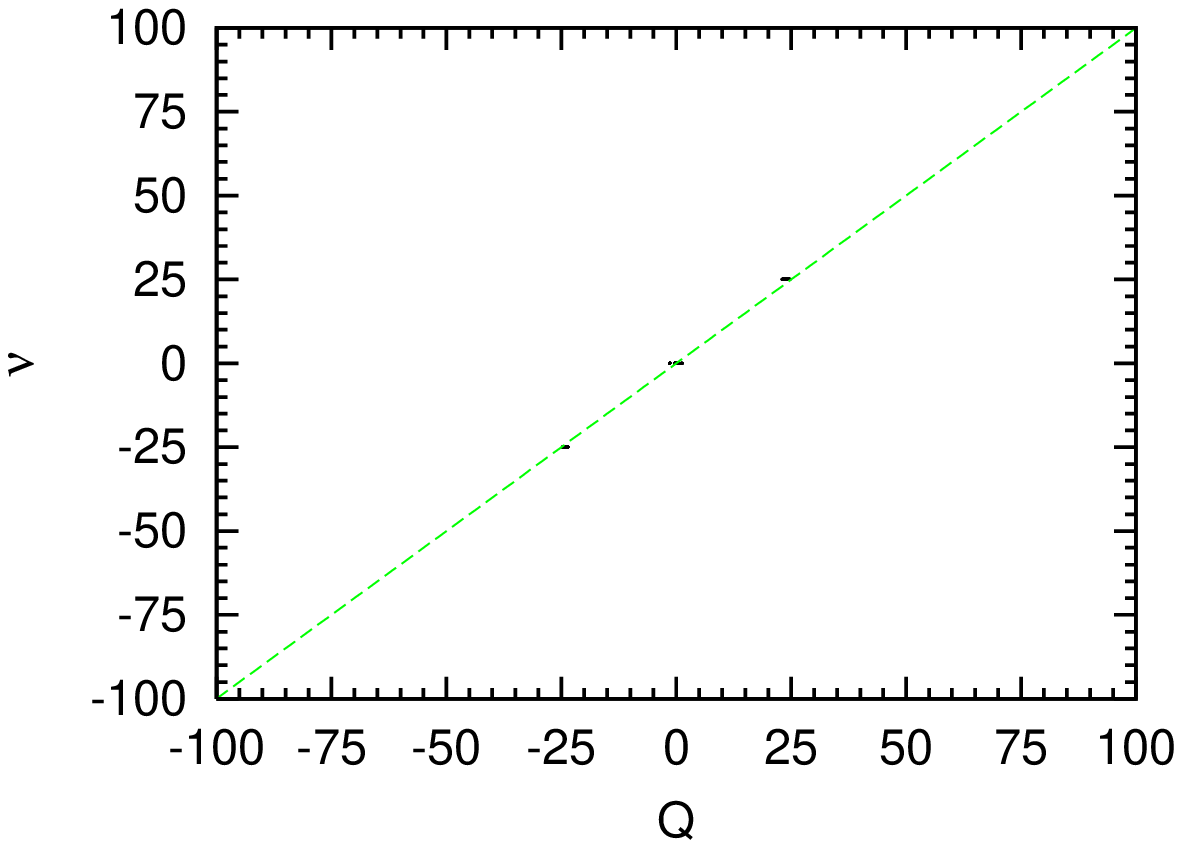,%
angle=270,width=7.4cm}
    \epsfig{file=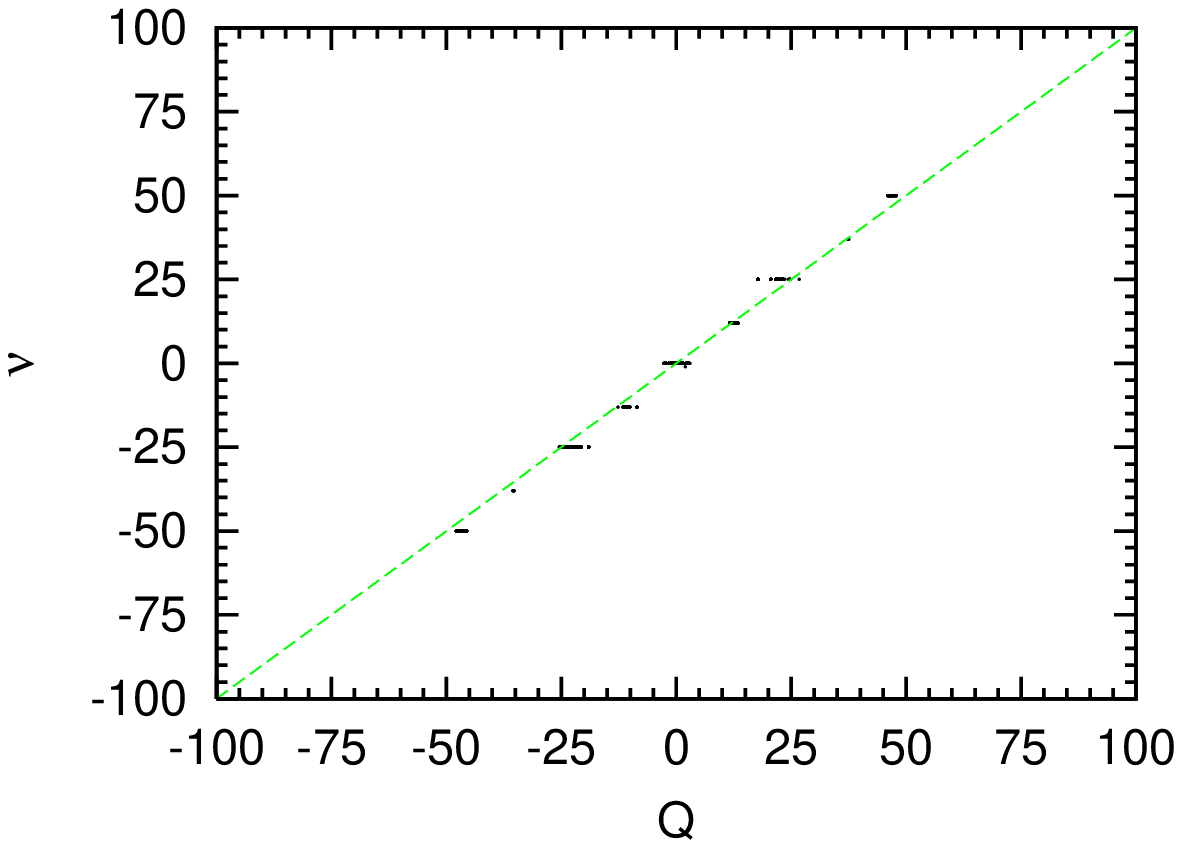,%
angle=270,width=7.4cm}
    \epsfig{file=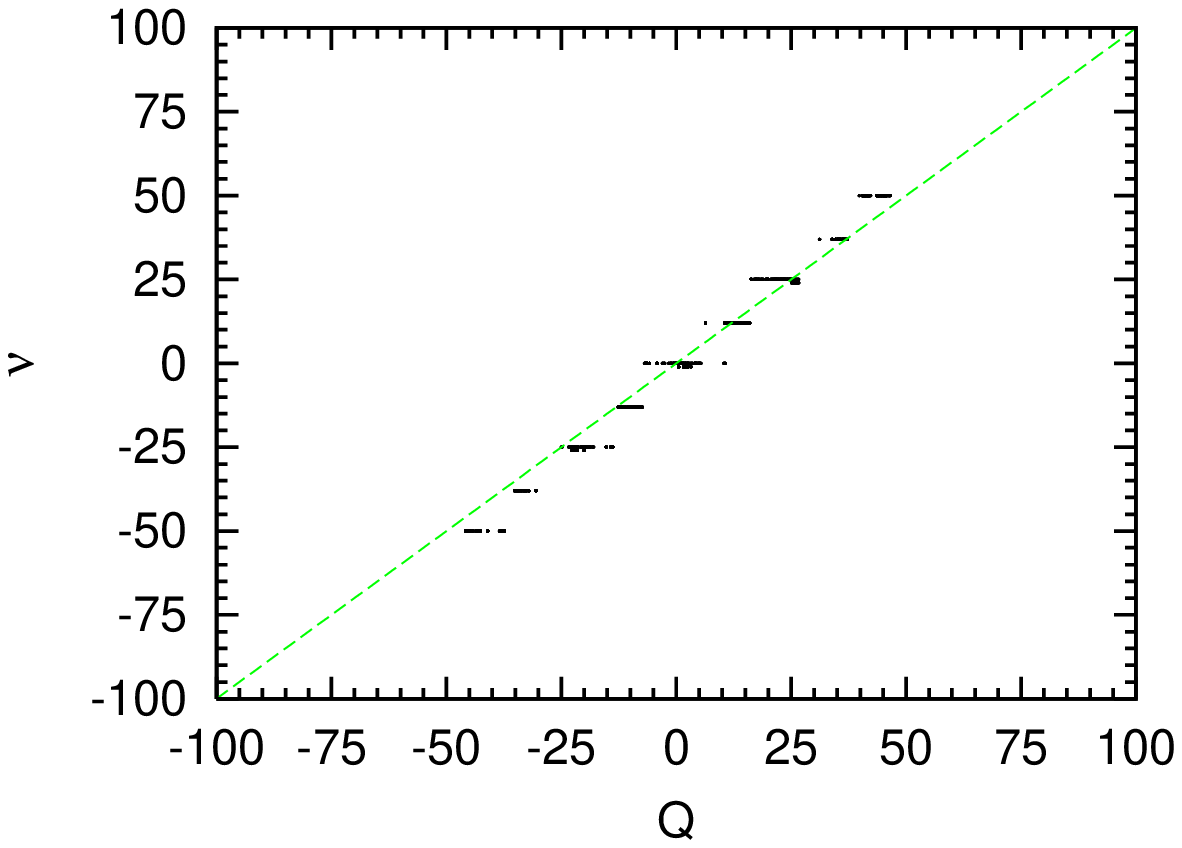,%
angle=270,width=7.4cm}
    \epsfig{file=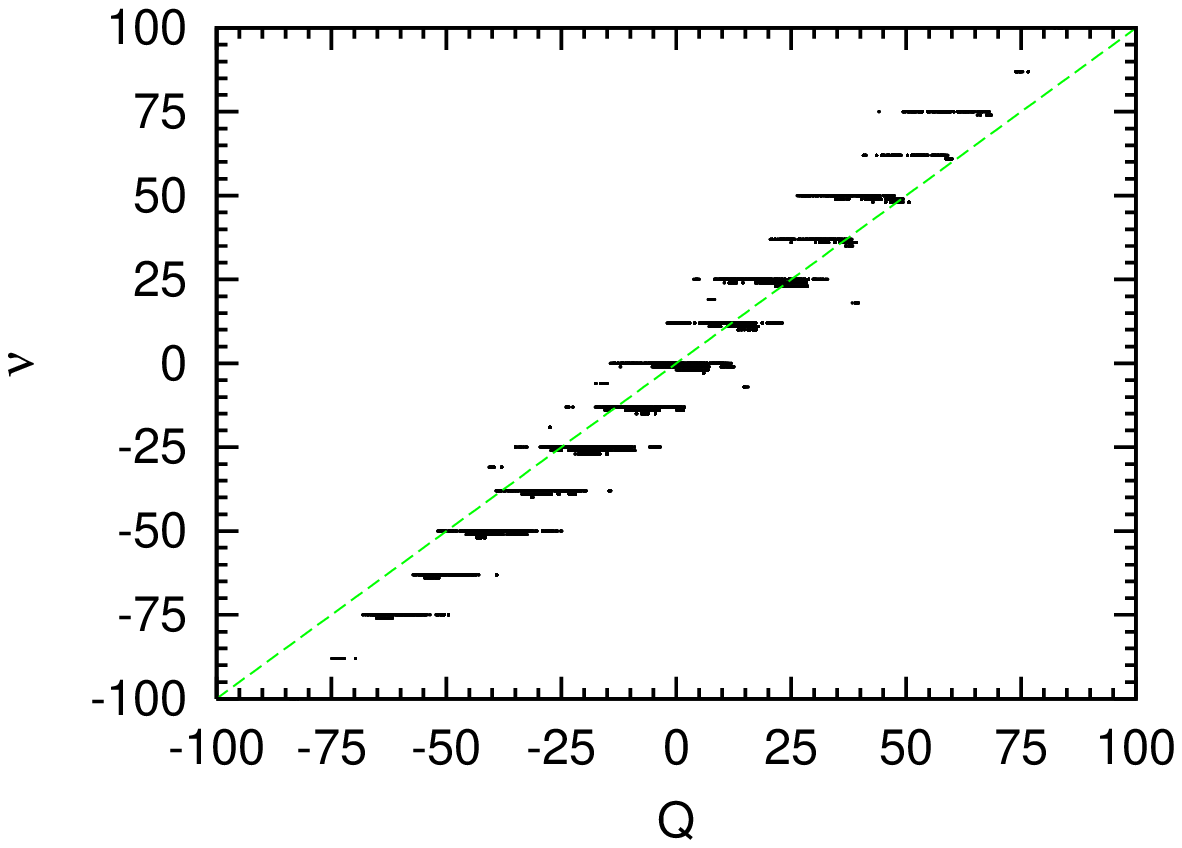,%
angle=270,width=7.4cm}
\caption{A scatter plot of the topological charge $Q$ (x-axis)
and the index $\nu$ (y-axis)
for solutions at $N=25$ with the action in the range
$S/\beta \le 100$ (top left), $100 \le S/\beta \le 200$ (top right)
$200 \le S/\beta \le 300$ (bottom left) 
and $300 \le S/\beta \le 600$ (bottom right).
}
\label{scatter-nu-q}
}

We also observe the accumulation of 
solutions with the topological charge close to
half-integer multiples of $N$.
The minimum action achieved by such solutions
increases linearly with $N$.
These are the solutions having one $1 \times 1 $ block
$\Gamma _\mu ^{(1)}=1$
with $n_1=1$ and $m_1=0$.
By choosing the $m_j$ ($j\ge 2$) so that
the arguments of the sine for the other blocks 
vanish in the large $N$ limit,
the topological charge (\ref{q-classical}) becomes
\beq
Q \simeq  N \left( \sum_j m_j  - M \right) + M +\frac{1}{2} \ ,
\label{q-half-integer}
\eeq
which coincides with the observed spectrum 
noting that $M = \frac{N+1}{2}$.
The action (\ref{action-classical}) is given by
$S/\beta \simeq 4 N$,
which nicely explains our observation from figure \ref{scatter-q-S}.

Next let us calculate the index
of the overlap 
Dirac operator
for the classical solutions,
and examine whether it agrees with the topological charge.
In figure \ref{scatter-nu-q} we present a
scatter plot of the topological charge $Q$ (x-axis)
and the index $\nu$ (y-axis)
for solutions at $N=25$ restricting the action
in four different regions.
%
%
We plot all the solutions in the displayed range 
without any restrictions.
For $S/\beta \le 100$, the index is
either $\nu = 0$ or $\nu = \pm N$, and the topological charge
turns out to be quite close to $\nu$, which nicely confirms
the index theorem.
For solutions with larger action, we observe the case
with $\nu$ close to half-integer multiples of $N$ 
in accord with (\ref{q-half-integer}).
While the index theorem is violated to some extent,
there still exists a strong correlation between $Q$ and $\nu$.
It is interesting that the smearing of the pattern occurs
mainly in the direction of $Q$. 
In this regard let us recall that the definition (\ref{def-q})
of $Q$ we have used is just a naive descritized version of the
continuum formula.
In order to recover an exact index theorem in the 
discretized setting, one may have to use a more sophisticated 
definition as in the commutative case \cite{Hasenfratzindex}.
Whether this is possible or not is an interesting open question.

\section{Configurations with the index of order 1}
\label{section:interpolating}

In the previous section we observed
that the topological charge 
(\ref{q-classical}) and the index
take only multiple integers of $N$ for classical solutions
with small action.
This is in striking contrast to the corresponding commutative theory,
where they take arbitrary integers, as we will discuss in
section \ref{section:commutative}.
In order to clarify the situation,
let us
construct configurations with the index of order 1
by interpolating the classical solutions in different topological sectors.
This can be achieved by replacing the integer parameters $m_j$
in the explicit form of the 
classical solutions
(\ref{explicit-Gamma}) by real parameters.
As the simplest case, 
we
consider 
the solutions (\ref{general_sol}) with $k=1$,
and generalize them to a one-parameter family
of configurations as
\beq
V_1=
\begin{pmatrix}
0&1& & &0\cr &0&1& & \cr& 
&\ddots&\ddots& \cr& & &\ddots&1\cr 1& &
& &0\cr
\end{pmatrix}  
~~~~~~,~~~~~~
V_2=
\begin{pmatrix}
1& & & & \cr & e^{2\pi i\mu/N}& & & \cr&
& e^{4\pi i\mu/N}& & \cr& & &\ddots& \cr & & & & e^{2\pi i(N-1)\mu/N}\cr
\end{pmatrix}   \ ,
\label{interpol-config}
\eeq
where $\mu$ is a real parameter. 
Since $\mu = M$ gives the absolute minimum of the action,
it is convenient to define $x \defeq \mu - M$.
As a function of $x$,
the action and the topological charge can be evaluated as
\beqa
\label{action-par}
S(x) &=& 4 N \beta
\left[ (N-1) \sin ^2  \frac{\pi x }{N}  +
\sin ^2 \left\{\pi \left( -1 + \frac{1}{N}\right) 
x  \right\} \right]   \ , \\
Q(x) &=& \frac{N}{2\pi} 
\left[ (N-1) \sin  \frac{2\pi x}{N}  +
\sin \left\{ 2 \pi \left( - 1 + \frac{1}{N}\right) 
x  \right\} \right] \ .
\label{q-par}
\eeqa

    \FIGURE{
    \epsfig{file=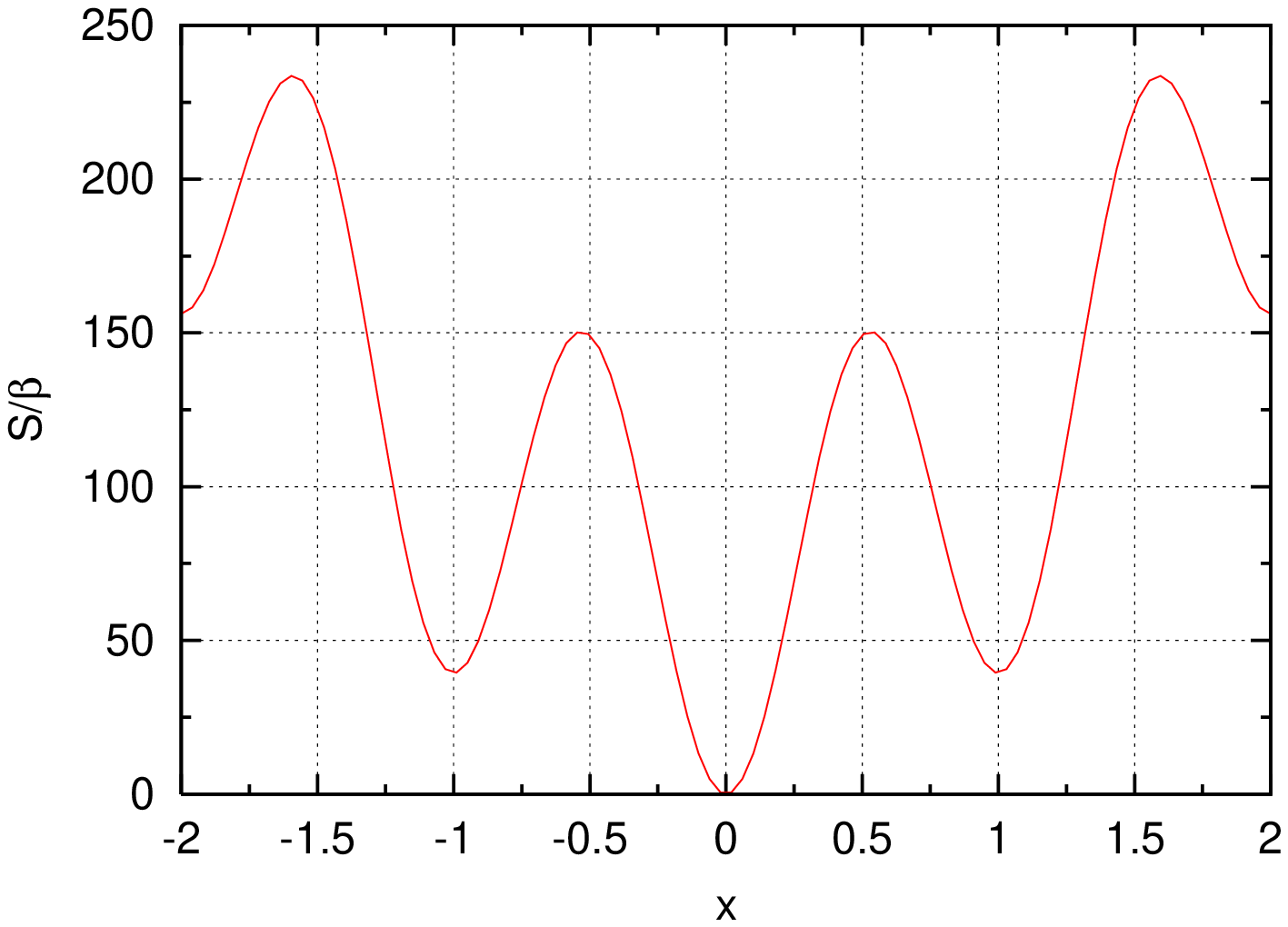,%
angle=270,width=7.4cm}
    \epsfig{file=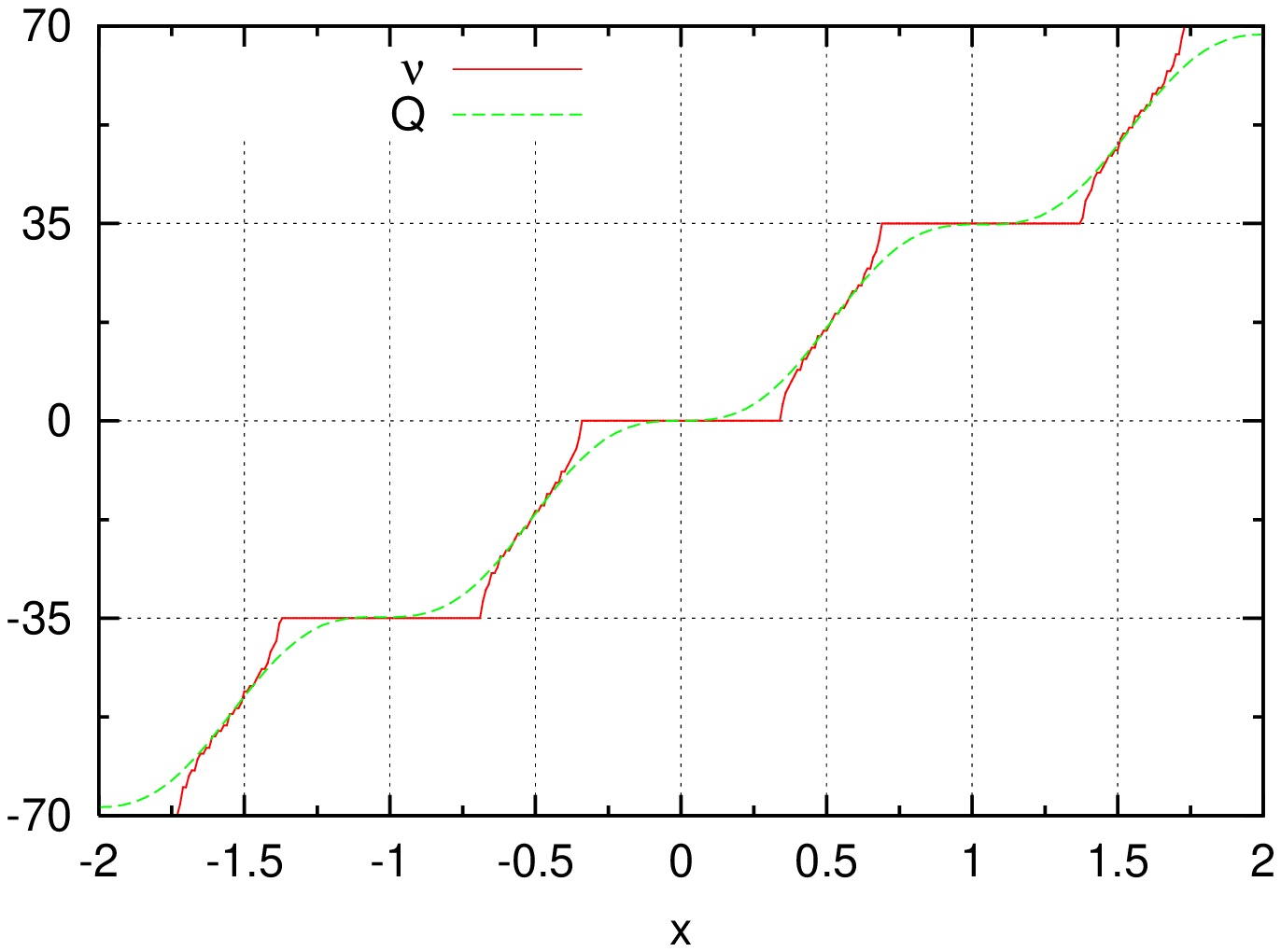,%
angle=270,width=7.4cm}
\caption{(Left) The action (\ref{action-par}) is 
plotted as a function of $x$
for $N=35$.
(Right) The topological charge $Q$ in (\ref{q-par})
and the index $\nu$ of the overlap 
Dirac operator
are plotted as a function of $x$ 
for $N=35$.
}
\label{fig:interpolating}
}

In figure \ref{fig:interpolating} (left)
we plot the action $S(x)$ against $x$ for $N=35$.
Let $n$ be the integer which is closest to $x$,
and consider the case where $|x-n| \sim O(N^{-p})$
with $p \ge 0$. Then, at large $N$, 
the leading contribution is given by
\beq
S(x)/\beta \simeq
\left\{ 
\begin{array}{lll}
4 \pi ^2 n^2  & \sim {\rm O}(1) & \mbox{~for~} p > \frac{1}{2}
\\
4 N \sin ^2 (\pi x)  & \sim {\rm O}(N^{1-2p})
& \mbox{~for~} 0 \le p < \frac{1}{2} \ .
\end{array}
\right. 
\label{action-largeN}
\eeq

In figure \ref{fig:interpolating} (right)
we plot 
the index $\nu$ of the overlap 
Dirac operator
and the topological charge $Q(x)$
against $x$ for $N=35$.
As we increase $x$ from 0 to 1, the index $\nu$ takes various integer
values between 0 and $N$. In this way we are able to construct
explicit configurations with $\nu$ of order 1.
We have also studied $N=15,25$,
and find that the result of the index $\nu$ is quite stable.
For instance, the region of $x$ which gives $\nu = 0$ is
$|x| < 0.36$ for $N=15$, and $|x| < 0.34$ for $N=25,35$.
This implies, 
in view of (\ref{action-largeN}),
that the
configurations with the index of order 1 constructed above
has an action of order $N$.

We also observe in figure \ref{fig:interpolating} (right)
that the topological charge does not agree with the index
for arbitrary $x$.
When $x$ is small, we obtain $Q \sim \frac{2}{3} N\pi^2 x^3$.
Therefore, in order to obtain $Q$ of order 1, we need to have
$x \sim {\rm O}(N^{-1/3})$, 
for which the action becomes of order $N^{1/3}$ due to 
eq.\ (\ref{action-largeN}).
However, for such a small $x$, the index is
zero in the large $N$ limit according to our discussion
in the previous paragraph.
This shows that the index theorem is violated even in the large
$N$ limit if the action is 
as large as O($N^{1/3}$)\footnote{We note that 
the admissibility condition
derived in ref.\ \cite{Nagao:2005st} allows configurations
with an action of O($N^2$). 
}.
It is of course possible that
the upper bound on $S/\beta$ for which the index theorem
holds {\em in general}
is actually less than O($N^{1/3}$),
say O(1).
In fact $Q$ and $\nu$ agree when $x$ is close to a half integer,
for which $S/\beta$ becomes of order $N$.
We consider this as accidental,
however, given the discrepancies observed for configurations
with smaller action.

Incidentally, we note that the configurations
at $x \sim n + \frac{1}{2}$ ($n \in \Z$),
which gives the local maxima of the action $S(x)$,
are closely related to the classical solutions
with the topological charge 
close to half-integer multiples of $N$
discussed in the previous section.
Indeed, for the two types of configurations,
the topological charge 
as well as the action coincides\footnote{Note, however, 
that they are not exactly equivalent configurations,
since the eigenvalue spectrum of $V_\mu$ is different,
and the former type of configuration is actually not a classical solution.
}
at large $N$.
In view of this, it is very likely that 
the classical solutions with the topological charge 
close to half-integer multiples of $N$
actually correspond to saddle-point configurations, 
instead of being local minima of the action.
These solutions are reminiscent of the
sphaleron configurations \cite{sphaleron}.

%


\section{Relationship to the commutative case}
\label{section:commutative}

In this section we review some known
results in the commutative case and discuss their
relationship to our results.
The commutative counterpart of our theory
can be obtained from (\ref{lat-action})
by replacing the star-product with the ordinary product.
The classical solutions are given by configurations
with a uniform field strength.
Explicitly, such a configuration can be constructed as
\beqa
U_1 (x) &=&
\left\{ 
\begin{array}{ll}
1 & \mbox{~if~} x_1\neq a(N-1) \nonumber \\
\exp(- 2 \pi i x_2 \tilde{Q}/ aN) & \mbox{~if~} x_1 = a(N-1)  \ ,
\end{array}
\right. \\
U_2 (x) &=& 
\exp(2 \pi i x_1 \tilde{Q}/ a N^2) \ ,
\label{commutative-sol}
\eeqa
where $\tilde{Q}$ is an integer, which corresponds to
the topological charge\footnote{A definition of 
the topological charge in the commutative case
can be obtained from (\ref{def-q}) by simply replacing the
star-product by the ordinary product.
This definition gives a non-integer value at finite lattice
spacing, and approaches the correct integer value $\tilde{Q}$ only
in the continuum limit.
However, 
there is a simple
geometric construction of the topological charge, which
gives an integer value even at finite lattice spacing.
This definition is used, for instance, in ref.\ \cite{GHL}.
}.
%
In fact one may obtain other solutions by
$U_\mu (x) \rightarrow e^{2\pi i h_\mu /N} U_\mu (x)$.
Since configurations obtained in this way with $h_\mu$
differing by integers are related with each other
by a large gauge transformation,
the gauge inequivalent solutions are obtained by restricting $h_\mu$
within the range $-\frac{1}{2} \le h_\mu < \frac{1}{2}$. 
Up to this degeneracy, which corresponds to the moduli space, 
there is essentially one classical solution in each topological sector
labeled by an integer $\tilde{Q}$.

For $x_1 = a(N-1)$, $U_1 (x)$ goes around the unit circle
in the complex plane
$\tilde{Q}$ times when $x_2$ goes from 0 to $a(N-1)$.
This implies that the configuration is singular\footnote{Note 
also that $U_2(x)\sim \exp (2\pi i \tilde{Q}/N)$ for $x_1 = a(N-1)$,
while $U_2(x)=1$ for $x_1 = 0$.
Therefore, $U_2(x)$ becomes singular when 
$0 < \frac{\tilde{Q}}{N} < 1$ in the continuum limit.
This singularity disappears, however, when $\tilde{Q}$ is kept
finite in the continuum limit or when $\tilde{Q}$ is a multiple
integer of $N$.
}
in the continuum limit with finite $\tilde{Q}$
since $U_1 (x)=1$ for $x_1 \neq a(N-1)$.
(Note, however, that the configuration is {\em physically}
smooth since the field strength is constant.)
This singularity disappears if and only if $\tilde{Q}$
is a multiple integer of $N$. In that case, the configuration itself
becomes totally smooth, and moreover it becomes translationally invariant
in the direction 2. For such configurations, the star-product
reduces to the ordinary product due to the definition (\ref{def-starprod}).
Therefore, the configurations satisfy the star-unitarity condition
(\ref{star-unitary2}), which
implies that they can be thought of as 
configurations on the discretized 2d NC torus.
In fact one can easily show that 
they correspond to the classical solutions 
(\ref{general_sol}) given by a 
single block ($k=1$)\footnote{
These configurations
have been studied earlier in refs.\ \cite{Giusti:2001ta,Kiskis:2002gr}
in the context of gauge theory in commutative space-time.}.
Note, however, that
there are many other classical solutions with larger action
in each topological sector
on the discretized 2d NC torus.




In the commutative case,
the probability distribution of the topological sectors, which 
are labeled by the index $\nu$,
can be calculated exactly in the continuum\footnote{We thank 
Hidenori Fukaya for clarification on this point.},
and it turns out to be 
$P(\nu) \propto e^{-S(\nu)}$, where $S(\nu)$ is the
minimum action in the topological sector $\nu$.
In terms of the lattice parameters,
$S(\nu)$ may be written as\footnote{For studies 
of the probability distribution $P(\nu)$
on the lattice, see refs.\ \cite{GHL,Bardeen:1998eq}.
}
\beq
S(\nu)=\frac{4 \pi^2 \beta}{N^2} \nu^2
\label{commutative}
\eeq
at large $N$ and $\beta$.
Since $\beta \propto \frac{1}{a^2}$ in the continuum limit,
the distribution scales as a function of $\nu/\ell$,
where $\ell=Na$ is the physical extent of the space.
Note that the probability for obtaining $\nu \lesssim {\rm O}(\ell)$
remains finite.

In the NC case, the classical solutions with
the action which is less than of order $N$ 
exist only in the topological
sectors labeled by $\nu$ which is a multiple of $N$.
The minimum action (\ref{action-singleblock})
for classical solutions in these topological sectors
agrees with (\ref{commutative}) for the reason explained above.
In the continuum limit, however, one has to take the
$a \rightarrow 0$ limit in such a way that
$\vartheta$ given by (\ref{theta-def}) is fixed.
Since $\beta$ should be sent to infinity as 
$\beta \propto \frac{1}{a^2}\propto  N$, which follows from
the scaling behavior of the correlation functions \cite{2dU1},
we obtain finite action only for $\nu =0$.
This suggests that
the probability of obtaining non-zero $\nu$ vanishes
in the continuum limit, which is consistent with the instanton calculus
in the continuum theory \cite{Paniak:2002fi}.
There the partition function has been written
as a sum over all the instanton configurations
with the total topological charge constrained to be 
equal to the magnetic flux, which is zero in the present case.

\section{Summary and discussions}
\label{summary}

In this paper we have studied
the index of the overlap
Dirac operator
in finite NC geometry, and clarified its basic properties
including the index theorem.
%
%
Our results confirm that
the overlap Dirac operator indeed
captures the topological nature of gauge theory 
in finite NC geometry, as in commutative lattice gauge theories.
An analytic proof of the index theorem 
extending the works \cite{Adams} in the commutative case
would be an interesting future direction.

%
In fact we have observed
a remarkable impact of NC geometry
on the topological properties of the theory.
As is well known, we encounter novel topological objects,
which are represented by infinitely many classical solutions in 
each topological sector.
However, we also observe the opposite effects.
The classical solutions with
an action less than of order $N$
should have an index $\nu$ which is a multiple integer of $N$.
%
While we were able to construct configurations
with the index $\nu$ of order 1
explicitly by interpolating the classical solutions, 
they have an action of order $N$.
The classical solutions with $\nu = \pm N , \pm 2N, \cdots$
have an action of order 1, but since it is strictly 
positive and proportional to $\beta$,
the action becomes infinite 
when one takes the $\beta \rightarrow \infty$ limit.
Thus we are left with the $\nu = 0$ sector 
in the continuum limit\footnote{Repeating our analysis 
in the case of finite torus would be 
straightforward, but we consider that the topologically nontrivial 
configurations would be even more difficult to survive the
continuum limit.}.
Confirmation of this statement in the full quantum theory
based on Monte Carlo simulation is reported
in a separate paper \cite{in-prep}.
%
%

The model we studied is the U(1) gauge theory
on a discretized 2d NC torus, whose
commutative counterpart has been studied extensively
in the literature for the reason that
it shares many dynamical properties with 4d
non-abelian gauge theories. 
The conclusion that the path integral is dominated
by the topologically trivial sector implies that
the $\theta$-term\footnote{The $\theta$ parameter, 
which appears here, 
represents the coefficient of the instanton number, 
and it should not be confused
with the $\vartheta$ representing the noncommutativity of the space-time.}
is irrelevant unlike in the commutative case \cite{FukayaOnogi}.
It would be interesting to investigate 
whether the suppression of non-zero indices is a general feature
of gauge theories on NC geometry, which is independent of 
the space-time dimensionality, the gauge group, the matter content 
and so on.
If the same property holds for the NC version of the
standard model, it suggests an exciting possibility 
that the strong CP problem is naturally solved due to the 
effects of NC geometry. 

Note, however, that
4d gauge theories in NC geometry
has problems of its own. 
Unlike the 2d case studied here,
the perturbative vacuum in the 4d case actually has
tachyonic instability due to the UV/IR mixing 
\cite{LLT,ruiz,Bassetto:2001vf,rf:MVR,rf:AL,Guralnik:2002ru}.
The system stabilizes by ``tachyon condensation'',
and finds a stable nonperturbative vacuum
\cite{4dU1},
in which the Wilson line corresponding to the tachyonic mode
acquires a vacuum expectation value.
Alternatively, one can stabilize the perturbative vacuum
by introducing an appropriate UV cutoff.
Although we do not know precisely
how we should construct a realistic model at this moment,
it is tempting to speculate that 
the strong CP problem may somehow be related to the physics
of string theory origin.

\acknowledgments

It is our pleasure to thank Hidenori Fukaya,
Satoshi Iso, Hikaru Kawai, Yusuke Kimura, Takeshi Morita,
Kenji Ogawa and Kentaroh Yoshida for valuable discussions.

\end{document}